\documentclass{article}
\usepackage[a4paper, total={6.5in,9in}]{geometry}
\usepackage[utf8]{inputenc}
\usepackage{amsmath}
\usepackage{amssymb}
\usepackage{graphicx}
\usepackage{authblk}
\usepackage{url}
\usepackage{natbib}
\usepackage{algorithm2e}
\usepackage{appendix}
\usepackage{rotating}
\usepackage{appendix}

\usepackage{longtable}

\usepackage[english]{babel}
\usepackage[autostyle]{csquotes}
\usepackage{indentfirst}

\vspace{-40pt}
\title{An Economic Topology of the Brexit Vote}
\vspace{-30pt}
\author[1]{Pawe{\l} D{\l}otko\thanks{P.D{\l}otko acknowledges support from the Engineering and Physical Sciences Research Council [grant number EP/R018472/1] as well as the support of the Dioscuri program initiated by the Max Planck Society, jointly managed by the National Science Centre (Poland), and mutually funded by the Polish Ministry of Science and Higher Education and the German Federal Ministry of Education and Research.}}
\affil[1]{Dioscuri Centre in Toplogical Data Analysis, Warsaw, Poland}
\vspace{-40pt}
\vspace{-30pt}
\author[2]{Lucy Minford}
\vspace{-30pt}
\author[2]{Simon Rudkin \thanks{\textbf{Corresponding Author}. Full Address: Economics Department, School of Social Sciences, Swansea University, Bay Campus, Swansea, SA1 8EN, United Kingdom. Tel: +44 (0)1792 606325 Email:s.t.rudkin@swansea.ac.uk}}
\affil[2]{Economics Department, Swansea University, United Kingdom}
\vspace{-30pt}
\author[3]{Wanling Qiu\thanks{Present Address: Accounting and Finance Department, School of Management, Swansea University, Bay Campus, Swansea, SA1 8EN United Kingdom. Email:wanling.qiu@liverpool.ac.uk}}
\affil[3]{School of Management, University of Liverpool, United Kingdom}
\vspace{-20pt}
\date{25th September 2021}
\begin{document}

\maketitle
\begin{abstract}
A desire to understand the decision of the UK to leave the European Union, Brexit, in the referendum of June 2016 has continued to occupy academics, the media and politicians. Using topological data analysis ball mapper we extract information from multi-dimensional datasets gathered on Brexit voting and regional socio-economic characteristics. While we find broad patterns consistent with extant empirical work, we also evidence that support for Leave drew from a far more homogenous demographic than Remain. Obtaining votes from this concise set was more straightforward for Leave campaigners than was Remain’s task of mobilising a diverse group to oppose Brexit.
\end{abstract}

\maketitle

\section{Introduction}

June 23rd 2016 saw the United Kingdom vote by a margin of 52\% to 48\% to leave the European Union. Britain's decision to exit the EU has become known as `Brexit'. Many narrative explanations for the result have been posited, some involving the `left behind' signalling to metropolitan elites, or the growth of Euroscepticism, fuelled by austerity and the challenges emerging from the Global Financial Crisis or else more deeply rooted in changes wrought by globalisation and European market integration. These theories have endogeneities and overlaps, implying significant statistical challenges so that conclusions remain elusive. As UK political parties continue to reconfigure themselves in its aftermath, understanding the voting patterns underlying the 2016 referendum outcome remains as important as ever.

This paper considers how local socio-economic characteristics combine to explain constituency-level voting behaviour in the EU referendum, and how this relates to prior and subsequent electoral voting behaviour, 
using topological data analysis (TDA).\footnote{To define a few key terms in this paper, `Leave' is a contraction of `leave the EU'; `Remain' that the UK should remain in the EU. In keeping with the literature, `Britain' and the UK are used interchangeably.} TDA is a data-driven approach that treats data as a point cloud and studies its topology. Born of work by \cite{carlsson2009topology}, TDA is well adopted in the physical sciences but is yet to take hold in the social sciences. It is free from assumptions about relationships and can capture coordinates on any number of axes to fine-grain all interactions. This approach is thus novel relative to the existing literature on the demographic characteristics underlying EU referendum voting patterns, which mostly assumes linear relationships. 

Using TDA, we explore relations between local socio-economic conditions and the 2016 EU referendum outcome. Complex interactions are thus reviewed in a topologically faithful fashion that preserves every element of the underlying data. Through this lens appear new patterns alongside the principal linear relationships already documented in the literature. We show that Brexit-voting constituencies are concentrated within a small part of the multidimensional data cloud, while Remain was highly spread. The discussion covers regional patterns in the point cloud and moves on to consider election outcomes. We therefore address both the geographical discussion begun in \cite{harris2016voting} and the demographic explorations in \cite{becker2017voted} and others. 

What follows is based upon observed proportions within parliamentary constituencies, proportions which are not highly variant \citep{carl2019european}. Constituencies have the advantage that they can be linked directly to parliamentary election results that comment on voter allegiance. In order to represent the dataset in a readily interpretable manner, we use the TDA Ball Mapper (TDABM) algorithm of \cite{dlotko2019ball}. A brief exposition of the approach follows in Section \ref{sec:method},\footnote{A fuller discussion is available in the supplementary material} the intuition being that any multidimensional dataset can be visualised in two dimensions by considering the strength of all co-locations within the point cloud that represents that dataset. 

The remainder of the paper is as follows. Section \ref{sec:lit} provides a brief overview of existing empirical work relating socio-economic factors to the referendum vote. Section \ref{sec:data} presents data forming the axes of the point clouds in our analyses, considering from a univariate perspective how values differ in Leave and Remain constituencies. The TDABM algorithm is introduced in Section \ref{sec:method}, while Section \ref{sec:full} applies it in our research context, constructing and analysing a TDABM plot based on chosen socio-economic axes and coloured by average proportion of constituency voting to Leave the EU. Section \ref{sec:further} considers the redrawing of the UK political map in elections after Brexit. Section \ref{sec:conclude} discusses conclusions from the TDABM analysis in light of existing empirical work on the UK 2016 EU Referendum result.

\section{Socio-economic patterns underlying the Leave vote}
\label{sec:lit}

An early attempt to bring TDA to the study of voting behaviour by ~\cite{ox_shape_of_brexit} offered a geographical topology perspective. Using one-dimensional homology and network theory, her work locates geographical interconnections between Leave and Remain votes. The approach we develop in this paper considers multidimensional constituency characteristics, so more firmly aligning with literature on socio-economic conditions, demographics and the decision to Leave.

While early analysis of correlations in polling data indicated broad roles for age, race and region in driving the Leave vote (\cite{Ashcroft2016}), subsequent research has revealed underlying patterns of greater complexity. The 2016 referendum vote has now been analysed extensively in existing literature against the available datasets, using OLS estimation techniques and largely linear models. \cite{becker2017voted} identify four broad hypotheses proposed as key drivers of the Leave result: EU exposure (trade, immigration and transfers), austerity and public service provision, demography and education, and economic characteristics (economic structure, wages and unemployment).\footnote{For literature on EU exposure through trade and on immigration, see e.g. \cite{los2017mismatch}, \cite{Clarke2017} and \cite{GoodwinMil}. On austerity policies as a driver, \cite{Fetzer2019}. Earlier discussions of demography and education are found in \cite{clarke2016importance} and \cite{sampson2017brexit} along with the role of cultural values in \cite{arnorsson2018causes}.} Grouping local authority-level data for a large number of covariates by these four categories, they use a best subset selection procedure, finding a linear model with prediction accuracy of 65\%. They conclude that factors like education profiles, skill levels and measures of deprivation are more important predictors of the Leave vote than other covariates they examine, and our variable selection in Section \ref{sec:data} is guided by these findings. \cite{alabrese2019voted} adopt a similar approach to Becker et al (2017) using both individual- and region-level data, finding that demographic and employment characteristics have the greatest predictive power for the Leave vote, alongside significant geographical heterogeneity. This study also concludes that the ecological fallacy is not a concern in this context: empirical patterns that hold at the local authority level are borne out at the individual level.

However, some or all of the many explanatory factors proposed as drivers of the referendum outcome may interact in a nonlinear fashion. Studies premised on linear relationships or bivariate considerations may miss important elements of the story, or stories. Additionally, inherent multicollinearity in demographic data with spatial aggregation usually forces researchers to focus on just a subset of relevant variable categories. \cite{zhang2018new}, for example, contracts the explanatory variable set to just the percentage in the upper social classes, the percentage with degrees and the percentage unemployed within any given local authority area. Such modelling strategies enable OLS analysis but leave questions around omitted characteristics. TDA can avoid much compromise of this nature.

\cite{Goodwinjrf2016} conclude broadly that Brexit voters were the ‘left behind’, with poverty and educational inequality among the biggest factors. However, there are nuances. One interaction effect is highlighted between an individual’s education level and the general skill level of their community: graduates in low-average-skill communities were more likely to vote Leave than graduates from high-average-skill communities. \cite{antonucci2017malaise} find the negative overall correlation between education and the Leave vote is driven by the strong association between intermediate education levels and Leave; the relationship is nonlinear. This association is stronger in conjunction with perceived economic decline. Rather than the ‘left behind’, they attribute much of the Brexit vote to the ‘squeezed middle’, voters identifying as working class while holding middle class jobs. Their analysis again underlines the role of interactions among multiple characteristics, and this motivates our application of TDA in this paper. Becker et al (2017) also emphasise the need to see “whether salient factors reinforced each other”, though of course investigating multiple interactions is impossible in an OLS setup. In general, we note repeated observations in the empirical literature that interactions and non-linearities have much to say on the question of why Brexit happened. 

\cite{liberini2019brexit} investigate microeconometric predictors of anti-EU sentiment, finding that individuals’ feelings about their circumstances (e.g. income) mattered more in referendum voting than their actual circumstances. We include subjective wellbeing in our analysis (cf. Alabrese et al, 2019 who use life satisfaction measures). Liberini et al (2019) also show that measurable variables like age are important not only in themselves but in conjunction with other characteristics. Our study takes its cue from this emphasis on nonlinear interactions between multiple covariates. In particular, we explore how similarity among constituencies based on a particular combination of multiple characteristics (leading to what we could term ‘group belonging’) translates into patterns of Leave and Remain support in the 2016 referendum vote using TDABM. The formation of these groups is based on non-spatial characteristics and \textit{a priori}, there is no reason to find geographically close constituencies in the same group\footnote{Post industrial regions may have similar demographics but can be found in the North East, North West, Wales, Scotland and the West Midlands. Equally all of those regions contain growing cities like Manchester, Leeds, Cardiff, Glasgow and Birmingham where new industry flourishes.}. Our interest here is in constituency characteristics as a sub-region of the main understood regions of the UK.

In terms of policy implications, the patterns revealed in our study are interesting for analysing the Leave and Remain campaigns: both their respective advantages in the runup to the referendum, and the extent to which their performance may have resulted in successfully converting voters to their side. Evidence from political science suggests that campaign effects themselves depend on interactions with voters' socio-economic characteristics. For instance, de Vreese et al (2011) find an interaction effect with political education: positive frames tended to be more effective for more politically aware individuals in the context of Turkish membership of the EU.\footnote{See also Hobolt (2007) and Van Elsas et al (2015).}

\cite{goodwin2018} argue that high information asymmetry in the runup to the UK referendum created potential for significant campaign effects. This potential was increased by the absence of clear partisan cuing on Brexit from the major political parties, with both Labour and Conservative parties split on the issue.\footnote{For this reason we do not include party affiliation as axes in our main TDABM plot, though we provide sensitivity analysis with these variables included. Our conclusions are robust, perhaps because - to the extent that it was informative on the Brexit vote - party affiliation tends itself to be a product of the socio-economic characteristics already accounted for in the analysis: see the supplementary material.} In an online survey experiment conducted in 2015, they examine the effectiveness of pro-EU and anti-EU frames. While receptivity to pro-EU frames is especially associated with certain characteristics (Labour support,\footnote{Analysis of TDABM plot coloured by political party affiliation available on request.} under 26's, and those undecided about the referendum), they find a significant interaction effect with education, with pro-EU arguments strongest among those with lower levels of education, though interactions with other socio-economic characteristics are not checked. The paper concludes on the failure of the Remain campaign to frame arguments adequately to persuade voters to their side.

Also interesting in this context are the findings of \cite{Shaw2017}, whose content analysis of nine TV debates in weeks preceding the referendum reveals core differences between the campaigns. They conclude that “Leave focused on a more consistent and tightly focused set of campaign themes, provided more explanation of those themes, and focused more on their own core issues than Remain” (p. 1020). They also analyse tactics employed in the debates by both sides, noting the tactic of tapping into emotion in particular.\footnote{p. 1027: “Both sides […] overtly pitched their argument to hit the emotions of individuals – less often geared towards a positive emotion.”} While they cannot capture the response of voters to emotive campaign messages, their premise is that voters could be influenced by the content of debates, placing emphasis on the campaigns themselves and their \textit{ex ante} ability to shift political opinion on the question of Brexit. On the other hand, microeconometric analysis of the result has tended to point to more fundamental drivers. Becker et al (2017 p.605) emphasize the explanatory role of “variables that seem hardly malleable in the short run by political choices (variables such as educational attainment, demography and industry structure).” The conclusion need not be that the campaigns were irrelevant to the outcome, however. The results of our topological analysis of socio-economic characteristics reveals clusters of `similar’ constituencies, and our suggestion is that these socio-economic clusterings are relevant for understanding how political campaign arguments and tactics are received, as well as for understanding constituencies' deeper predispositions on the question.

\section{Data}
\label{sec:data}

As a base for analysis we use parliamentary constituency data as compiled by Professor Pippa Norris for work in \cite{thorsen2017uk}.\footnote{Data accessed from \url{https://www.pippanorris.com/data}.} Demographic data from the 2011 census is merged in the \cite{thorsen2017uk} set, permitting analysis of the socio-demographic space upon which the Brexit vote played out. Because constituencies do not correlate directly with counting districts used in the referendum, \cite{hanretty2017areal} constructs estimates of the percentage of voters for Leave and Remain in each constituency. We combine UK general election results for 2015 and 2017 from the downloaded data with 2019 election results, allowing consideration of the EU referendum in the context of changing party affiliation.\footnote{Election data downloaded from \texttt{https://commonslibrary.parliament.uk/research-briefings/cbp-8647/{\#}fullreport}.} 2015 represents the last election before the referendum and indicates prior political leanings of constituencies.

\begin{table}
	\begin{center}
		\begin{scriptsize}
			\caption{Summary Statistics and Univariate Tests}
			\label{tab:sumstats_reduced}
			\begin{tabular}{l l   c c c c c c c c }
				
				\hline
				Question & Variable & Mean & s.d. & Min & Max & \multicolumn{3}{c}{Leave v Remain} & Corr\\
				& && & & & Leave & Remain & Diff & \\
				\hline

				2015 Vote (\%)& Labour & 32.35&16.5&4.51&81.3&31.65&33.58&-1.93 & -0.06\\
				& Conservatives  &36.66&16.16&4.67&65.88&39.6&31.46&8.14*** & 0.24\\
				& Liberal Democrats  &7.82&8.36&0.75&51.49&6.96&9.34&-2.38** & -0.26\\
				& Others &23.17&11.85&6.09&65.33&21.78&25.62&-3.83** &\\
		       Housing Tenure & 1: Owned &64.05&11.42&20.48&85.68&67.04&58.78&8.26***&0.44\\
& 2: Social rental&17.99&7.8&4.59&50.63&16.73&20.21&-3.48***&-0.23\\
&3: Private rental&15.9&6.41&5.55&42.1&14.26&18.8&-4.54***&-0.47\\
&4: Other &2.06&0.65&0.82&7.93&1.97&2.21&-0.24***&-0.22\\
Household Composition&1: Married&33.33&5.76&14.63&46.33&34.43&31.39&3.04***&0.35\\
&2: Cohabit&9.74&1.49&3.5&13.82&10.05&9.19&0.86***&0.25\\
&3: Other &45.91&7.35&32.26&71.85&43.82&49.61&-5.78***&-0.48\\
Car Ownership & 0 Cars&25.54&11.57&7.86&66.7&22.84&30.32&-7.49***&-0.40\\
&1 Cars&42.3&2.99&28.24&50.25&42.78&41.46&1.32***&0.31\\
&2+ Cars&32.16&10.9&4.38&57.96&34.38&28.22&6.17***&0.34\\
NSSEC&1: (see notes) &40.04&8.6&20.25&64.58&38.28&43.16&-4.88***&-0.39\\
&2: &19.99&2.76&10.59&26.88&20.99&18.22&2.77***&0.61\\
&3: &25.82&6.47&9.73&40.31&28.26&21.51&6.74***&0.65\\
&4:&5.42&2.87&1.62&22.59&5.27&5.69&-0.42&-0.05\\
Qualifications & 1: None + L1&47.49&4.01&35.22&60.4&45.51&50.99&-5.48***&-0.78\\
&2: L2 + apprenticeship&33.62&2.13&26.81&38.38&34.45&32.15&2.3***&0.59\\
&3: L3 + L4&18.89&3.35&10.96&29.38&20.04&16.86&3.18***&0.56\\
Self-reported Health & Very Good &37.61&8.44&15.64&67&39.55&34.18&5.37***&0.47\\
& Good&18.61&3.38&8&23.8&20.43&15.38&5.04***&0.78\\
&Other&38.81&8.75&21.31&67.68&35.07&45.42&-10.36***&-0.74\\
Deprivation&0&42.14&6.98&22.21&59.71&41.39&43.47&-2.08***&-0.22\\
 & 1 &32.56&1.75&28.14&38.38&32.66&32.37&0.3&0.09\\
 & 2 or Higher&25.3&6.29&12.15&44.77&25.95&24.16&1.78**&0.21\\
Age &1: <18&21.1&2.39&12.54&33.79&21.41&20.56&0.85***&0.25\\
&2: 18-24 &9.3&3.59&5.73&32.68&8.5&10.7&-2.2***&-0.35\\
&3: 25-59&46.65&3.41&36.49&63.23&45.74&48.26&-2.53***&-0.48\\
&4: 60+&22.95&5.34&7.78&40.56&24.35&20.47&3.88***&0.43\\
				\hline
			\end{tabular}
		\end{scriptsize}
	\end{center}
	\raggedright
	\footnotesize{Notes: Variables organised by question; total for each constituency on each question is 100\%. All variables from 2011 Census except 2015 vote percentages from \cite{thorsen2017uk}. Categories combined from individual answers from census data. Owned housing tenure combines owned outright and mortgage categories; `other' combines shared and rent free categories. `Other' household composition includes living alone, lone parent, all-student households and all others. NSSEC (National Statistics Socio-Economic Classification) categories are 1) higher managerial + higher professional +lower manager + small employer; 2) intermediate + lower supervisor; 3) semi-routine + routine; 4) never worked + long-term unemployed. `Corr' provides the correlation between the proportion of individuals responding to the 2011 Census as belonging to a category and the \cite{hanretty2017areal} estimated Leave percentage. There is no correlation value for others owing to the diverse range of parties within the category. Leave v Remain divides constituencies on whether \cite{hanretty2017areal}-estimated leave percentage is greater (smaller) than 50\%. Difference augmented by significance of two-sample t-test for equality of means: * - 5\%, ** - 1\% and *** - 0.1\%.}
\end{table}

  The constituency characteristics we investigate are housing tenure and occupancy, motor vehicle access, NSSEC status, qualification levels and self-reported health. Our dataset contains information for 45 different categories within these seven questions reported from the 2011 census.\footnote{Full summary statistics for each of the 45 categories are available in the supplementary material.} The assumption is that constituencies' demographic makeup did not vary greatly between 2011 and 2016. While question-by-question analysis may be interesting, the 2016 referendum results are the consequence of all factors in combination. TDABM is a big data approach and can readily extend to 45 axes without undue concern for degrees of freedom. However, categories for which many constituencies register very low proportions often create connections, as the balls encompass their full range long before the other characteristics\footnote{To understand the intuition for this consider the tenure variable percentage living rent-free. This has a minimum value close to 0\% and a maximum close to 4\%. A radius of 2 could then include all values of rent-free living iff the values on every other axis were the same. This extends then to cases where the radius is much larger, such as the 23 used in this paper, where the whole range of rent free can be covered and still allow variation in other axes.}. One treatment is to normalise all variables onto the range $\left[0,1\right]$, but this then gives equal importance to all variables in the plot. Since all values in this paper are on the scale 0-100\% we do not rescale here. Instead, we have slightly reduced the number of axis variables by merging some categories where there is a rationale to do so. For instance, in any constituency a very small proportion of households own 3 or 4 cars. We consider the proportion of households with \textit{2 or more} cars on the basis that this is the more salient information. We also merged categories which tend to be distributed in similar areas of the BM plot (e.g. NSSEC categories and self-reported health; see the supplementary material for BM plots coloured by axis variable). Certain categories making up very small proportions for all constituencies are dropped, e.g. the proportion of households for which all household residents are 65+. Age categories are included as separate axis variables. 
  
  The 27 axis variables employed in the analysis are listed in Table \ref{tab:sumstats_reduced} along with descriptive statistics for the average proportion of each category in each constituency. These provide a flavour for the data and permit some basic testing of links with Brexit voting. Using two-sample t-tests we report the difference between the values of each characteristic in Remain and Leave areas, reporting a positive value in the `Diff' column whenever the average value amongst Leave is higher.
 
  Across the dataset, broad averages are consistent with correlations reported in existing literature on the Brexit vote: lower education levels, higher levels of deprivation, lower-skilled occupation and poorer health are all associated with the Leave vote, supporting an exclusion story whereby those who felt socio-economically `left behind' drove the referendum result (e.g. \cite{GoodHeath2016b}, \cite{Mckenzie2016}, \cite{hobolt2016brexit}, \cite{inglehart2016trump}, \cite{bromley2018brexit}). Existing work has verified that data at higher levels of aggregation is representative of the individual data for these demographic variables and their Brexit-voting correlations (Alabrese et al, 2019).

\section{Methodology}
\label{sec:method}
TDA views data as a point cloud, with the position of each data point within the cloud defined by its values on each of the axes that comprise the cloud. Each variable in the data set may become an axis, subject to the requirement that it is ordinal and that the realised values within the dataset have sufficient variation. In this paper the cloud comprises the 27 characteristics of constituencies outlined in Table \ref{tab:sumstats_reduced}. When there are only two dimensions that point cloud is analogous to a scatter plot. As the dimensionality increases, a visualisation tool becomes necessary to obtain the valuable inference that is offered by scatter plots. TDABM addresses this, providing an abstract representation of the point cloud that maintains full connection to the underlying dataset.\footnote{A full exposition of the methodology and discussion of evaluating outcomes is provided in the supplementary material.} 

First, the TDABM algorithm selects a point at random from the dataset and constructs a ball of radius $\epsilon$ around it. In two dimensions a ball is simply a circle, but TDABM operates in any number of dimensions. The initial selected data point is the first landmark. Any other data points within the ball are considered to be covered by the ball. TDABM will then select a second landmark from the uncovered set, marking as covered any points within a ball of radius $\epsilon$ surrounding that landmark. Continuing to iterate point selection, the algorithm finishes when there are no uncovered points.

Relative positioning of the balls is obtained through the presence of points in the intersection of two balls. Where there is a non-empty intersection, an edge is drawn between the two landmarks. Density of the cloud is captured by resizing the representation of the ball to reflect the number of points it contains; larger balls signify more points within radius $\epsilon$ of the landmark. Summary statistics may be produced for each ball including the average value of the outcome of interest for the points within the ball. In the visualisation, balls are coloured according to a function of the points they contain. Primarily below this means colouring according to the average \cite{hanretty2017areal}-estimated Leave percentage. We also tell a voting story by colouring by outcomes of the general elections of 2015, 2017 and 2019. 

Outcomes from the TDABM algorithm are dependent solely on the choice of the radius $\epsilon$. No hard rule exists about the optimal choice, but it is straightforward to iterate over radii to verify the robustness of conclusions. Outcomes are also dependent on the random landmark selection. However, as shown in subsequent sections, the broad inference of TDABM is consistent over multiple applications of the algorithm with different random seeds. Visualisations can thus be understood with confidence and bootstrapped confidence intervals constructed on any metric derived from the TDABM plot.

In what follows $\epsilon=23$ is used; a demonstration of the strong robustness of the key messages in this paper to $\epsilon$ selection is provided in the supplementary material.

\section{Results}
\label{sec:full}

\subsection{Ball Mapper Results} 

\begin{figure}
	\begin{center}
		\caption{Leave Vote Percentages ($\epsilon=23$)}
		\label{fig:reduced}
		\includegraphics[width=15cm]{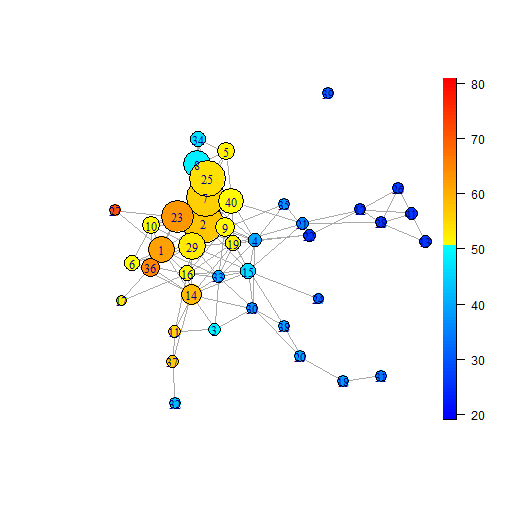} 
	
	\end{center}
	\raggedright
	\footnotesize{Notes: TDA Ball Mapper diagram using reduced set of Census 2011 variable categories. Details of the combination of categories are provided in the data section. Colouration is by \cite{hanretty2017areal} estimated leave percentages with the 50\% cut off being towards the upper end of the green shading. Axes are constructed from the combination of categories within questions. Data from \cite{thorsen2017uk}. Construction of diagram using R package \textit{BallMapper} \citep{dlotko2019R}.}
\end{figure}

Figure \ref{fig:reduced} provides a TDABM graph with $\epsilon=23$ showing a large concentration of balls to the centre left with three arms extending towards the bottom and right. Colouration is by average Brexit support in each ball. Each ball is a collection of constituencies with broadly similar characteristics from our combined category set. A join between two balls means that there is at least one constituency sitting in both balls. As this plot seeks to represent 27 dimensions in two dimensional form, there is no direct interpretation of the horizontal or vertical direction. The TDABM graph allows us to see the shape of the data; to find out more about specific variables' behaviour we would colour by that variable. The supplementary material provides plots coloured by each axis variable; plots in Figures \ref{fig:reduced} and 3-\ref{fig:full} are coloured by non-axis variables which can be thought of loosely as `outcome' variables.

\begin{table}
    \begin{center}
        \caption{Leave Vote Percentages Summary ($\epsilon = 23$)}
        \label{tab:colour}
        \begin{tabular}{c c c c c c c c c c c c c c c }
        \hline
             Ball & Size & Leave && Ball & Size & Leave && Ball & Size & Leave && Ball & Size & Leave \\
            \hline
            1&98&61.57&&12&7&26.34&&23&135&62.29&&34&28&46.81\\
            2&188&57.00&&13&12&26.72&&24&2&32.27&&35&4&37.63\\
            3&11&49.31&&14&61&58.16&&25&158&54.21&&36&50&64.33\\
            4&16&39.81&&15&30&45.74&&26&3&27.66&&37&9&56.51\\
            5&43&52.20&&16&29&51.48&&27&3&70.49&&38&6&39.29\\
            6&31&51.11&&17&2&53.32&&28&7&26.7&&39&3&30.46\\
            7&168&55.68&&18&5&37.94&&29&103&52.87&&40&92&51.03\\
            8&104&48.45&&19&27&51.99&&30&8&37.67&&41&12&26.54\\
            9&55&50.94&&20&7&39.35&&31&4&34.59&&&\\
            10&39&51.81&&21&9&35.27&&32&3&41.79&&&\\
            11&12&54.93&&22&9&26.56&&33&13&40.57&&&\\
            \hline
        \end{tabular}
    \end{center}
\footnotesize{Notes: Ball numbers related to the topological data analysis ball mapper plot of our reduced category dataset with $\epsilon=23$ plotted in Figure \ref{fig:reduced}. Size is the number of constituencies contained within the ball. Leave is the average \cite{hanretty2017areal} estimated Leave percentage for the constituencies contained within the ball. These Leave values correspond to the colouring of the balls in Figure \ref{fig:reduced}.}
\end{table}

Three points emerge immediately from Figure \ref{fig:reduced}. First, the comparative concentration of constituencies within the upper left of the plot. Though this has no direct interpretation in terms of the values of the explanatory variables, it does inform that the constituencies here are very similar to each other in all dimensions. There is only one disconnected ball, informing that most constituencies have at least one other to which they are quite similar. Second,  the Leave vote, coloured on the yellow to red scale, is concentrated within a core part of the space to the left of the plot. Leave particularly covers the larger balls in the upper left. Finally, we note that the Remain colouration, on the blue scale, sits away from the main mass of the plot and is more thinly spread. This tells us there is greater heterogeneity between Remain voting constituencies. We return to this important observation subsequently.

Table \ref{tab:colour} provides average Remain percentages and numbers of constituencies for each of the balls in the diagram. Together with Figure \ref{fig:reduced}, the recurring message is that there are more Remain-supporting balls than Leave: that is, there are more balls where the average Hanretty-estimated Leave percentage is lower than 50\%. Furthermore, of the 11 balls containing more than 50 constituencies only one, ball 8, has an average estimated Leave percentage below 50\%. Discounting ball 8, the average number of constituencies in a ball with less than 50\% estimated Leave vote is just 9.5\footnote{Including ball 8 the average rises to 13.8.}. By contrast the average size of a ball with greater than 50\% estimated Leave percentage is 68.6 constituencies. These statistics reinforce the message from viewing the TDABM plot.

TDABM output retains the data points in each ball for interrogation. Ball 32, to the bottom centre of the shape, contains Glasgow East, Glasgow North East and Glasgow South West. It is understood that Scotland was pro-Remain and these seats are held by the pro-EU Scottish National Party. Glasgow South West provides a characteristic link into many other similar constituencies from other industrial cities in ball 37 which features areas of Liverpool like Bootle, Walton and West Derby. Blackley and Broughton in Manchester, Birmingham Erdington and Nottingham North are also in ball 37. We may view this arm as being the more deprived areas of major cities. Aside from Glasgow this explains the pro-Brexit colouration (\cite{GoodHeath2016b}). We refer to this string of balls as `arm A' of the plot. 

The string of Remain balls running through 38, 20, 18 and 31 contains more cities. Ball 38 has Nottingham East and Nottingham South, Sheffield Central and Newcastle upon Tyne East. These are Labour seats that were held in 2019. The major difference with Ball 32 comes from the high proportion of highly qualified individuals in Ball 38. Deprivation in ball 38 is also much lower than ball 32. Ball 20 contains Nottingham East as a bridge. It has a higher level of deprivation than ball 38, but maintains the high levels of residents with post-compulsory education. Moving along the arm (`arm B') into 18 and 31 we see the NSSEC levels of the jobs move lower and the levels of deprivation rise. Ball 18 picks up Manchester Central, Leeds Central, Tottenham and West Ham. Ball 31 is then Glasgow Central, Leeds Central, Manchester Central and Liverpool Riverside. These are very different communities to those of ball 32, with ball 31 having more young adults, higher incidences of private rental, higher qualification levels and more residents whose job roles are either intermediate or lower supervisor on the NSSEC classification. The balls are very similar in having low car ownership, low self-reported health and higher proportions in the household composition group that combines those living alone, lone parents and households of students\footnote{The dominant group within this category is students in ball 31, and lone resident and lone parent in ball 32. However, the differences are small in absolute terms because of the low proportion of households in each set.}. 

Finally the strong Remain arm (`arm C'), heading out from balls 4 and 21, features constituencies such as Bristol West, Manchester Withington, Hove and Brighton Pavillion in Ball 22. Ball 13 is entirely London boroughs and includes Islington North, Hackney, Bethnal Green and Bow and Hammersmith. Ball 26 has the Cities of London and Westminster, Kensington and Hampsted and Kilburn. There are obvious differentials between these more affluent areas of London and the boroughs of ball 13. Kensington forms the overlap with Ball 41. Also in 41 are Hackney North, Vauxhall, Lewisham and Deptford, Hammersmith and Islington North. These are all Labour seats and within London. We may understand the differences between balls here through deprivation, qualifications and the extent of social rent. Moving to the right of this arm we see increasing deprivation and reduced car ownership.

Finally, we note Ball 39 as an outlier Remain ball. This ball contains Richmond Park, Twickenham and Wimbledon. These are all suburbs of West London with high levels of home ownership, high qualifications and low deprivation. The age distribution is more skewed towards older residents and occupations tend to be from higher NSSEC groups.

These three arms are all very different, hence we do not see any connectivity between them. There are the more deprived constituencies of Glasgow in arm A, the diverse regional city centres in arm B, and the London boroughs in arm C. Whilst the Remain arms beg analysis for their differences, there is a converse similarity about the Leave voting areas, the oranges and yellows on Figure \ref{fig:reduced}. Towards the top of the main body of balls are balls 5 and 25 with Hanretty-estimated Leave percentages of 52\% and 54\% respectively. In the centre left are 23, 1 and 36 that are deeper orange in colouration and have Hanretty-estimated Leave percentages of 62\%, 62\% and 64\% respectively. These are very strongly Leave-supporting balls. Finally there is the yellow colouration stretching down into arm A, balls 11 and 37 having Hanretty-estimated Leave percentages of 55\% and 57\%, respectively.

Balls 25 and 5 are predominantly rural and the Hanretty-estimated Leave votes are very similar. Here we find constituencies like the Derbyshire Dales, Central Devon, South Suffolk and West Worcestershire; also towns like Aylesbury, Newark, Shipley and Stratford-on-Avon, which all had marginal votes to Leave. As may be understood with the overlap to Ball 8 there are constituencies like North Somerset, Monmouth and Horsham whose vote was marginally in favour of Remain that also sit in this ball. These are areas where home ownership is high, having 2 or more cars is common and the highest NSSEC occupations are found. Qualifications are high and self-reported health in this area of the plot is very good. Models have aligned many of these characteristics with Remain, but as we see the overall combination leans to Leave. Regions of the TDABM plot like this highlight the importance of interactions within the data. Visualisation facilitates a reappraisal of the understood relationships from additively separable regressions.

Balls 1 and 36 are where some of the strongest average Leave vote is found. Here the constituencies are predominantly urban, with many linked to industrial decline. Here we find Blackburn, Burnley and Bradford of the traditional textile towns; also former coal-mining areas such as Merthyr Tydfil, the constituency of Normanton, Pontefract and Castleford, Rhondda and Bolsover; and the former steel areas of Scunthorpe, Redcar, Stocksbridge and Rotherham. These are constituencies where health is poorer, qualifications lower and deprivation is high. However, there are similarities with balls 5 and 25 too. In balls 1 and 36 we see high home ownership, marriage and similar prevalence of 1 car households and upper-middle NSSEC class occupations.

Moving down into arm A and towards ball 32 we find balls 11 and 37, containing deprived suburbs and a mixture of voting behaviours. In ball 11 sit Gateshead, Leeds East and Birmingham Erdington, all with Hanretty-estimated Leave percentages above 55\%. Here too sit Edmonton and Newcastle-upon-Tyne Central with Leave percentages below 50\%. In ball 37 are Bootle, Middlesbrough and Nottingham North, all with very high Leave percentages. There is then Glasgow South West which serves as the link into the blue coloured ball 32. These constituencies have lower home ownership, the dominant category being social rent. There are lower levels of marriage and more households without access to a car. Health and qualifications are lower and deprivation higher. 

\subsection{Remain Heterogeneity, Leave Concentration}

Primary inference from Figure \ref{fig:reduced} is that Leave support is far more concentrated within the space than Remain. The scale on the right reporting Leave percentages as estimated by \cite{hanretty2017areal} places 50\% between the light and dark blues; all Leave constituencies are in the centre of the big shape in the left part of the plot. It is also immediate that the biggest balls correspond to those voting to leave the EU, while those wishing to remain are more spread out. The strongest Remain constituencies are found on arm C, extending out to the right, though each arm goes to Leave percentages of less than 40\%. There are more marginal Remain areas to the top and left of the main connected shape. That Brexit-favouring constituencies are more jointly similar on these axes than others comes through strongly in the plot. 

\begin{figure}
    \begin{center}
        \caption{Brexit Constituency Distribution Robustness}
        \label{fig:bmrange}
        \begin{tabular}{c c}
            \includegraphics[width=7cm]{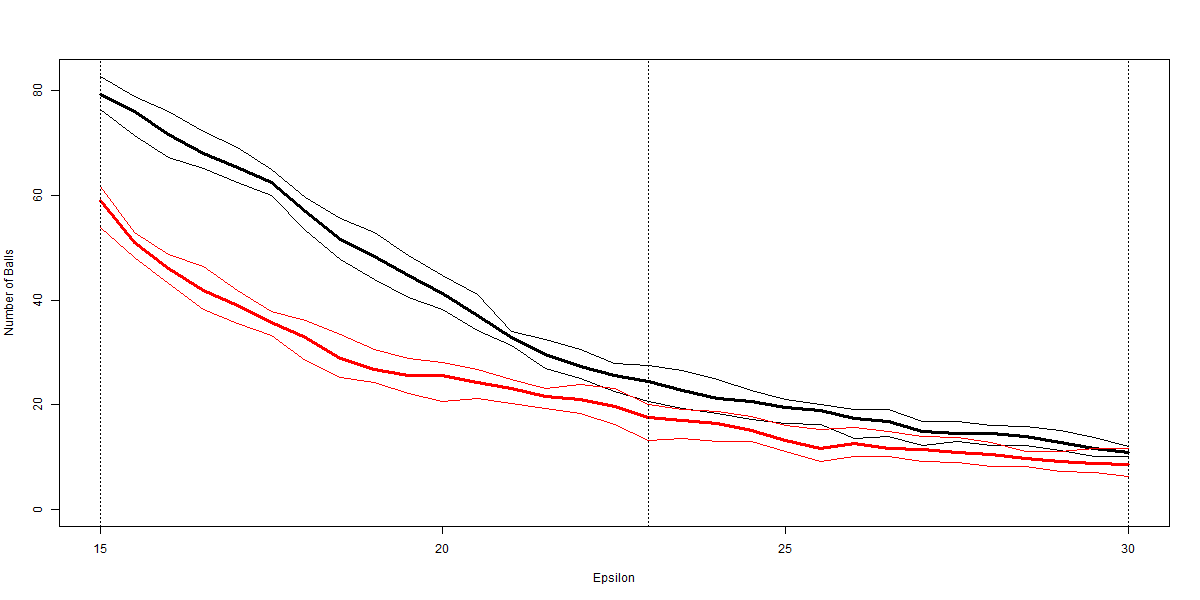}
            & \includegraphics[width=7cm]{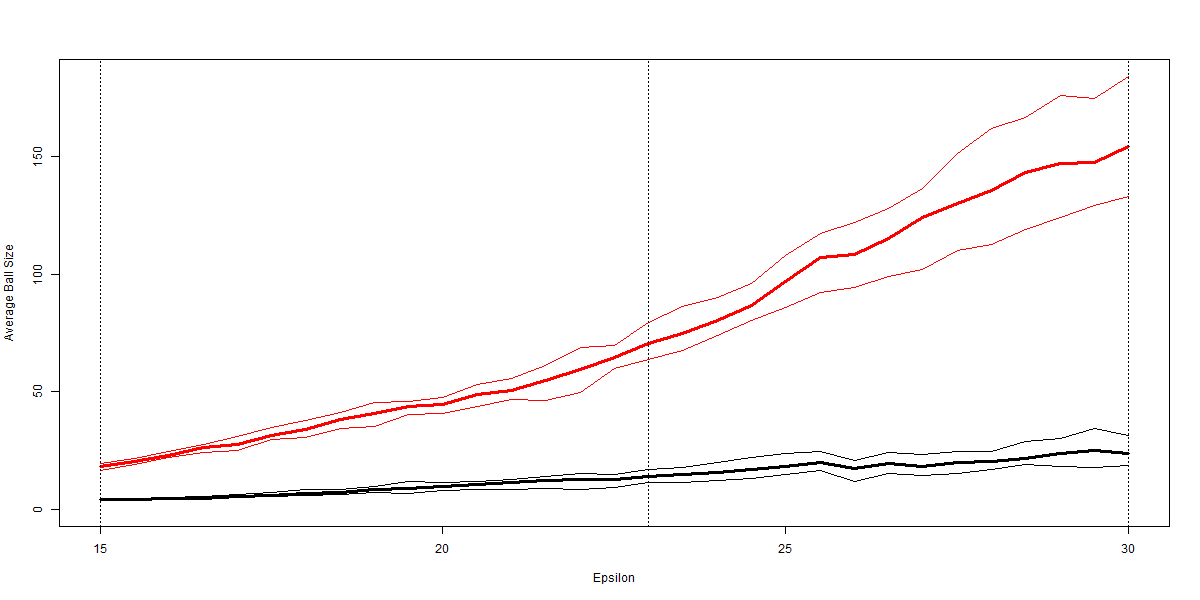}\\
            (a) Number of Balls  &  (b) Average Ball Size \\
        \end{tabular}
    \end{center}
\footnotesize{Notes: Figures plot the average ball size and number of balls which satisfy the condition that the average Leave percentage for the ball is less than 50\% and greater than 50\%. These averages are based on 10000 repetitions of the topological data analysis ball mapper algorithm at radii $\epsilon \in \left[10,30\right]$ in increments of 1. Lines relating to Remain balls are plotted in black, whilst lines relating to Leave balls are plotted in red. 95\% confidence intervals from the 10000 estimates are plotted as thinner lines in the respective colours.} 
\end{figure}

We investigate the robustness of Remain heterogeneity versus Leave concentration using the BM algorithm. By iterating the BM algorithm 10000 times over radii in the range $\epsilon \in \left[10,30\right]$ we can understand more about the nature of Leave and Remain balls. From each iteration we collect the average size of balls that have a colouration value less than 50\%. We also collect the number of balls that average a Remain vote from each BM graph. Figure \ref{fig:bmrange} shows the results.

Figure \ref{fig:bmrange} has two panels. In each panel the black lines relate to the Remain balls, whilst the red lines relate to those balls with an average \cite{hanretty2017areal} estimated Leave percentage above 50\%. The left panel reports that the number of Remain supporting balls is consistently higher than the number of Leave supporting balls. Likewise the size of the Remain supporting balls is much smaller than the average size of the Leave supporting balls. Confidence intervals from the 10000 repetitions confirm these results to be significant. Consequently our illustration in Figure \ref{fig:reduced} is no exception in showing the Remain concentration.

Turning this towards an understanding of the Remain campaign's failure, we focus on voters in marginal areas. Marginal areas are coloured in light blue in Figure \ref{fig:reduced}: a line of these cuts through the plot from ball 3 to ball 35 through 33 and 4. There is then a second marginal pair in balls 8 and 34.  The prominent Remain areas are then outside the cut. We could ask: what campaign message could have converted more voters in those light yellow-coloured balls, like 6 and 40, to vote Remain? Our contention is that, if campaign messages are more successful when targeted to demographic characteristics, a message targeting ball 6 would not simultaneously convert voters in ball 40 (while retaining core supporters in ball 30). What might be effective in mobilising votes in constituencies in one part of the space may not be effective elsewhere. This diversity necessitates different messages and opens potential for conflicting signals that diminish impact. The conclusions of \cite{Shaw2017} regarding the relative `incoherence' of the Remain campaign are then less surprising.

Analysis of the TDABM results pointed to the clear heterogeneity of the Remain voting clusters. The variation from Glasgow suburbs to the centres of the major cities and to the boroughs of London could not be more stark. Not only are the geographic and political differences clear, the overall difference across all characteristics taken together is also large. As the outcome of the referendum is based on the total nationwide vote seeking extra support within any constituency has merit. Balls indicate where marginal constituencies sit in the characteristic space. 

As an example, let us consider ball 19 which connects to balls 4, 15 and 33 that were all coloured blue, but is itself yellow. In ball 19 we find smaller urban areas like Cheltenham, Chester, Exeter and Hove which are all estimated to have voted Remain, but also other similar rural conurbations such as Colchester, Lincoln, Poole and Worcester that were all estimated to vote Leave. Moving up from ball 19 into balls 40 and 7 deprivation continues to fall, but moving left into balls 29 and 2 the proportion of households suffering two or more of the deprivation indicators rises. Ball 19 does not contain so-called ``red wall'' seats where the Leave campaign had strong appeal (\cite{harris2016voting}, \cite{antonucci2017malaise}, \cite{los2017mismatch}), rather this is a set of constituencies where Remain messages had chance to resonate. The plot therefore serves as a useful post-campaign evaluation tool.

\section{Further Analysis}
\label{sec:further}
Our results show the contrasting nature of Leave and Remain support, the former concentrated in a small area of the demographic characteristic space while the latter is highly spread; in other words, when all interactions among demographic variables are taken into account Leave-supporting constituencies are more alike than Remain constituencies. Colouring the plot by the 2015, 2017 and 2019 election results we use TDABM to illustrate how the changing political landscape plays out on our characteristic space. Secondly we evaluate the effect of using the full set of categories in the dataset. That is we use the elements that are combined into the axes described in Table \ref{tab:sumstats_reduced}\footnote{A table of summary statistics for each of the options within the Census 2011 questions is provided within the supplementary material.}. Each discussion demonstrates further the value of visualisations from the TDABM algorithm.

\subsection{Political Parties and Brexit}
\label{sec:pol}

\begin{figure}
    \begin{center}
        \caption{Conservative Gains from Labour at December 2019 ($\epsilon=23$)}
        \label{fig:cgfl}
        \begin{tabular}{c c}
            \includegraphics[width=7cm]{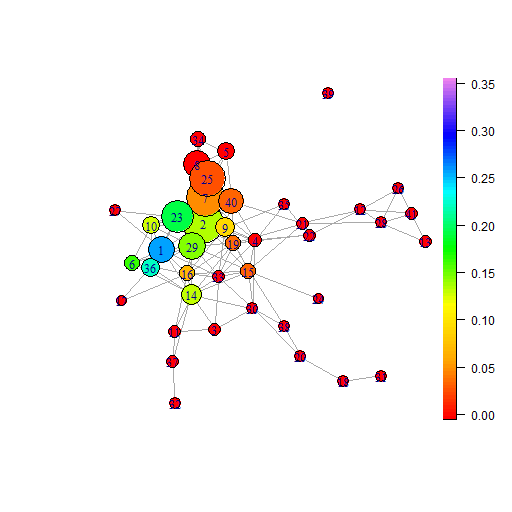}& 
             \includegraphics[width=7cm]{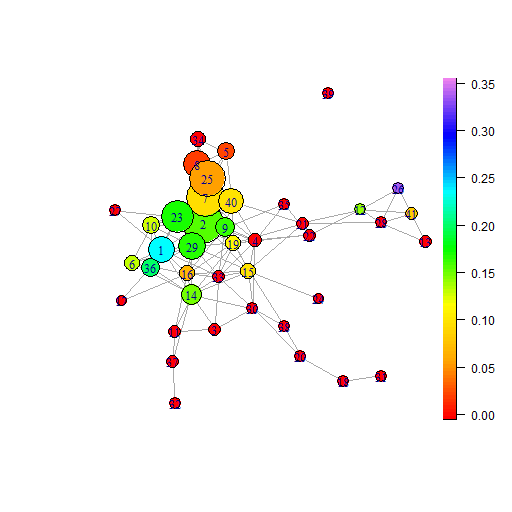}\\
             (a) Gains versus 2015 & (b) Gains versus 2017 \\ 
             \includegraphics[width=7cm]{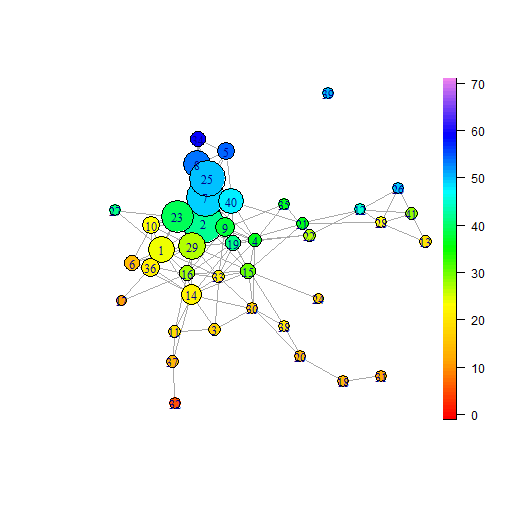} & 
             \includegraphics[width=7cm]{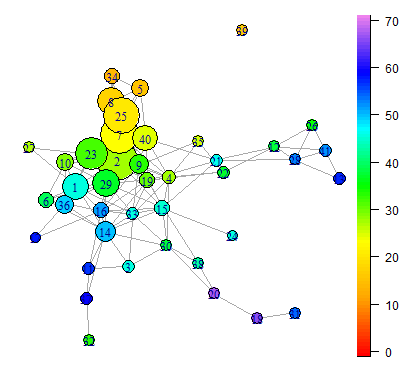} \\ 
             (c) Conservative vote 2015 & (d) Labour vote 2015\\
        \end{tabular}
    \end{center}
\raggedright
\footnotesize{Notes: Topological data analysis ball mapper diagram constructed using \textit{BallMapper} \citep{dlotko2019R}. Plots coloured by proportion of constituencies in each ball won by Conservative party from Labour in December 2019 General Election relative to stated past election. Colouration key provided in right-hand bar. Panels (c) and (d) are coloured by the 2015 vote shares for Conservatives and Labour respectively. Data from \cite{thorsen2017uk} and www.gov.uk. }
\end{figure}

Politics and the Brexit vote are intrinsically linked, the impasse in parliamentary proceedings at the time of the exit deal negotiations ultimately leading to a third general election within 5 years. Figure \ref{fig:cgfl} can help to visualise how exactly that 2019 election played out, and how its results link back to the Brexit question. Employing the same axes as the previous plots facilitates rapid comparison of election voting patterns and Brexit voting patterns. Panels (a) and (b) are coloured by the proportion of formerly Labour constituencies won by the Conservative party in December 2019. To add reference we also colour by the 2015 election results in panels (c) and (d).\footnote{Balls registering a low percentage in panels (a) and (b) could signify either `conservative throughout' or `not conservative throughout'. Further plots analysing support for other parties available on request.}

These plots tell a clear story, showing how long-standing political allegiances were disrupted by the referendum (\cite{Ashcroft2019}, \cite{Cooper21}). In terms of campaign emphasis, the Conservative Party message was a simple ``Get Brexit Done'' while Labour laid out a spending programme directed at remodeling society (\cite{Guard1}). Panels (a) and (b) show Labour losses to the Conservative Party located in areas of the plot where Brexit support was strongest in Figure 3. Ball 1 contains the post-industrial areas, particularly in Northern England and South Wales. Balls 2, 23 and 29 also have about 20\% of constituencies gained by the Conservatives, having had average Labour votes of more than 30\%. These balls include Sedgefield, which had been the seat of Labour Prime Minister Tony Blair but fell to the Conservatives in 2019.  Conservative gains versus both 2015 and 2017 are then concentrated in this part of the shape, not to the centre or right where the Leave vote was weaker, reiterating the centrality of the Brexit question to subsequent election outcomes, and indicating party repositioning from the Conservative party in response to the political shock of the EU referendum (\cite{Hayton2021}).

If the reader draws an imaginary line through the connected balls running from ball 37 through balls 11, 14, 19, 40, and 5, then the Leave-voting section of the plot is everything including and to the left of that line, with the exception of the majority Remain balls 8, 34 and 27 (cf. Figure \ref{fig:reduced}). The lower panels of Figure \ref{fig:cgfl} show that seats in this Leave-leaning area range from solid Conservative in the upper left corner (tending to be rural, e.g. South East Cornwall, Devizes) to large leads for Labour in the lower left of the shape in 2015 where several post-industrial and former mining constituencies cluster. In panels (a) and (b) there is then a general gradient sloping downward from a peak at ball 1, where the proportion of Labour seats falling to the Conservatives in 2019 passes 20\%. Comparing panels (a) and (b) to panel (c) reveals that in balls 7 and 2 the Conservatives added constituencies socio-economically similar to those already held in 2015, while the bigger proportions in balls 23, 36 and 1 show a Conservative swing in constituency clusters with characteristics traditionally associated with Labour (panel d), the phenomenon sometimes described as the collapse of the `red wall’ \citep{cutts2020brexit}. These balls contain traditionally Labour-voting constituencies in the north of England, North East Wales and the Midlands and the plots emphasize the centrality of Brexit to some of the Labour Party’s long-time faithful.\footnote{e.g. Walsall North, Mansfield, Great Grimsby and Stoke-on-Trent in ball 1, Wrexham in ball 2, Burnley in ball 36.} Commentary at the time pointed to the “increasingly unstable alliance of Labour’s left and centre, its remain and leave electorates, and its middle-class and working-class bases” (\cite{Guard2}), and panel (d) confirms that Labour party support in 2015 was much more spread out across the BM plot than Conservative support in panel (c). Some strong Labour support in 2015 is found in the two strongly Remain arms of the plot stretching to the right (arms B and C), backing up the general link between Labour party affiliation and propensity to vote Remain reported in literature (\cite{alabrese2019voted}, \cite{goodwin2018}). However, the TDABM analysis provided here highlights visually the types of constituencies deviating from this general correlation.

Panels (a) and (b) are generally similar, though Conservative gains relative to 2015 are more modest than versus 2017, the highest proportion in an individual ball being 23\% and 33\% respectively, reflecting growing popular frustration with parliamentary gridlock over Brexit (\cite{Ashcroft2019}, \cite{Cooper21}). Most constituencies that moved from Labour in 2017 to Conservative in 2019 were also Labour in 2015. A good counter-example is seen in Figure \ref{fig:cgfl} panel (b) where the highest proportion of Conservative gains relative to 2017 comes at the very right in ball 24. This contains Kensington, a seat Labour had taken from the Conservatives at the 2017 General Election. In panel (a) of Figure \ref{fig:cgfl} it is coloured red, as Kensington is not a gain versus the 2015 result.

\subsection{Full Dataset}

\begin{figure}
	\begin{center}
		\caption{Full Dataset: $\epsilon=23$}
		\label{fig:full}
		\includegraphics[width=15cm]{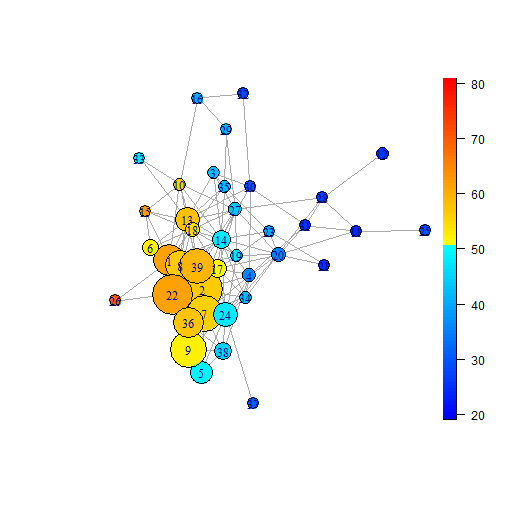} 
	\end{center}
	\raggedright
	\footnotesize{Notes: Topological data analysis ball mapper diagram constructed using \textit{BallMapper} \citep{dlotko2019R}. Colouration is by \cite{hanretty2017areal} estimated leave percentages with the 50\% cut off represented by the transition from lighter to darker colours. Data from \cite{thorsen2017uk}.}
\end{figure}

Section \ref{sec:data} discussed the decision to reduce the number of categories used as axes. This mitigated concerns over spurious connections among balls in the BM plot due to low registered proportions on certain categories shared by almost every constituency. To demonstrate consistency between plots based on the merged categories versus the full set of unmerged socio-economic categories, we turn briefly to the mapper plot with full set of axis variables. Figure \ref{fig:full} depicts a similar story of concentration of Brexit-voting constituencies within the plot to Figure \ref{fig:reduced}. For comparability with the main plots the same radius, $\epsilon=23$ is used. Colouration is also set so that there is a switch from a blue scale to a red scale at 51\%. At the core of the shape we see the strong oranges, the high Leave vote that determined the result. To the right of the plot sit all of the majority-Remain constituency groupings, the blue balls which are smaller and more numerous (as before). While there are indeed more interconnections between the Remain balls, the plot still conveys the impression of a dispersed periphery. Indeed, there is a strong likeness between the shape of the two plots.\footnote{Note that while the BM algorithm has assigned different numbers to the balls in Figure \ref{fig:full} relative to Figure \ref{fig:reduced}, the constituencies within the large, highly connected balls making up the Leave section of the plot are consistent with the earlier plot.} 

\section{Conclusions}
\label{sec:conclude} 

It is impressive that what Becker et al (2017, p.605) acknowledge to be “very simple empirical models” can explain a significant amount of variation in the Leave vote share across local authorities. However, these simple models employed elsewhere in the literature do omit nonlinearities recognised to be key. Here we have taken a different approach and instead use TDABM to investigate the clustering of observations in a point cloud constructed from a rich set of socio-economic covariates, going on to analyse how Leave and Remain support varies across that map. The primary emphasis with this method is not on individual covariates, or trying to posit a linear relationship between them and the Leave vote (on which there are plenty of existing contributions), but on whether constituencies share things ‘in common’ in terms of those covariates taken together. Once commonalities are revealed, the researcher can dig down into the groupings to see which covariates drive the pattern in different parts of the space. This is a novel way of approaching the referendum data which accounts for multiple variable interactions.

Using a constituency-level dataset, this paper has demonstrated that Leave-voting constituencies tend to share more commonalities than Remain-voting. That is not to say that all Leave-voting constituencies are alike - there is heterogeneity among them - but balls identified in Figure \ref{fig:reduced} as majority Leave-voting constituencies are larger, less numerous and more interconnected (indicating members common to more than one ball) than balls containing high proportions of Remain-voting constituencies. Those are small, more numerous and more spread out in the space. 

The TDABM plot established two distinct strings of strongly Remain-leaning groups: the diverse city centres in arm B, many large university student populations, contrasting with the string of London boroughs in arm C. Interestingly, further analysis of the balls shows how the combinations of shared characteristics change as we move along these arms from the centre towards the right: qualifications and NSSEC classifications fall and deprivation rises. Arm A highlights the Remain-voting constituencies of Glasgow in an outlying ball, linked on socio-economic characteristics to other deprived pro-Brexit constituencies of major cities. Meanwhile at the opposite end of the shape lie two groupings of affluent rural constituencies, sharing many characteristics with Leave-voting constituencies but nevertheless supporting Remain.

While, as with regression analysis, these results do not comment on the causal mechanisms driving the Leave vote, considering the data from a different angle can suggest new avenues for the modelling of those channels. When a subset of observations register similar values on a number of covariates then those observations can be said to form a group (a ball in the BM graph). Our results suggest that group (ball) membership may be directly relevant for members’ propensity to vote Leave or Remain. However, it may also be relevant for their propensity to be influenced by certain campaign messages or tactics on the Brexit question. This latter possibility casts the referendum campaigns as a channel linking socio-economic factors to referendum voting behaviour; socio-economic factors may affect how particular types of political messaging are targeted or received.

The suggestion we raise for future research is that the multidimensional clusterings of socio-economic characteristics shown in Figure \ are useful for understanding voting behaviour and perhaps, in particular, how political campaign messaging is more or less effective at mobilising voters. Figure \ref{fig:reduced} reveals where the Remain campaign did not convert opinion, though without directly commenting on the reasons for that. Our results point not so much to a failure by Remain as to the comparative simplicity of the task faced by the Leave campaign, catering to relatively similar groups of constituencies while Remain-voting constituencies were highly diverse. To convert more marginal constituencies to vote Remain without alienating core supporters would have required a more differentiated (and perhaps therefore less coherent) campaign; indeed, this may explain the relative incoherence of the Remain campaign noted by \cite{Shaw2017}. 

Many critiques of data-driven approaches abide, and variable choice is clearly of great importance. Axis variables selected here are ruled by the existing literature and the available data within the readily accessible dataset of \cite{thorsen2017uk}. However, the strength of the TDABM algorithm comes from the ability to deal in multiple dimensions. To that end the presentation here can be readily extended and an analysis of any ordinal constituency characteristic incorporated. Next logical steps would see the approach applied to individual level data where there is more heterogeneity and a further interest in the interactions of multiple characteristics.

We also provide further support in Section \ref{sec:pol} for Brexit as an external political shock and instigator of significant party change. The issue of Brexit is far from settled in the minds of a still-divided electorate \citep{Axe2021} and remains high on the political agenda; voter sentiment on the question shapes party manifestos and government policies today. Figure \ref{fig:cgfl} illustrates the role of the EU referendum in redrawing the UK political map, with political parties repositioning more or less successfully in light of changing patterns of allegiance among the electorate \citep{Cooper21}. It remains to be seen whether the coalitions built by the Conservative party in 2019 will hold into the next election.

\section*{Data Availability Statement}

The data that support the findings of this study are openly available from the website of Professor Pippa Norris at \texttt{https://www.pippanorris.com/data/} and from the United Kingdom Commons Library at \texttt{https://commonslibrary.parliament.uk/research-briefings/cbp-8647/{\#}fullreport}.

\bibliography{brexit}
\bibliographystyle{apalike}

\end{document}